\documentclass{ptephy_v1}
\usepackage{graphicx}
\usepackage{bm}
\begin{document}
\title{Role of small-norm components in extended random-phase approximation}
\author{Mitsuru Tohyama}

\address{{Kyorin University School of Medicine, Mitaka, Tokyo
181-8611, Japan}
\email{tohyama@ks.kyorin-u.ac.jp}}

\begin{abstract}
The role of the small-norm amplitudes in extended RPA theories such as  
the particle-particle and hole-hole components of one-body amplitudes and the two-body amplitudes other than two particle - two holes components are investigated for
the one-dimensional Hubbard model 
using an extended RPA derived from the time-dependent density-matrix theory. 
It is found that these amplitudes cannot be neglected in strongly interacting regions where the effects of ground-state correlations are significant.
\end{abstract}
\maketitle
\section{Introduction}
The random-phase approximation (RPA) is formulated based on the Hartree-Fock (HF) ground state where it is assumed that the lowest single-particle states (hole states) 
are completely occupied and other single-particle states (particle states) are empty. Therefore, RPA consists of particle (p)-hole (h) and h-p amplitudes.
When the effects of ground-state correlations are explicitly taken into account in extended RPA approaches such as the renormalized rRPA \cite{rowe1}, 
the self-consistent RPA (SCRPA)\cite{scrpa1,scrpa2},
the extended second RPA (ESRPA) \cite{srpa}, its response function version \cite{taka1}
and the small amplitude limit of the time-dependent density-matrix theory (TDDM)\cite{WC,GT,GT89}, 
the single-particles states become partially occupied. As a consequence, the
p-p and h-h amplitudes in addition to the p-h and h-p amplitudes should also be included as one-body transition amplitudes 
to satisfy sum rules and properties of spurious states \cite{adachi,spur1,spur2}. The two-body amplitudes other than 
the 2p-2h and 2h-2p amplitudes may also play some roles when the effects of ground-state correlations are included. 
In most applications, however, these additional amplitudes which have small eigenvalues of a norm matrix
have usually been neglected.
In this paper we investigate how the p-p and h-h amplitudes and the 2p-2p, 2h-2h and ph-ph amplitudes
are incorporated into physical states using an extended RPA approach (ERPA) based on the ground state in TDDM.
We use the one-dimensional Hubbard model at half filling  which in momentum space 
has an equal number of particle and hole states and thus allows us to investigate the effects of the above-mentioned additional amplitudes.
The paper is organized as follows the formulation is given in sect. 2, the results are shown in sect. 3 and sect. 4 is devoted to summary.

\section{Formulation} 
We consider the Hamiltonian $H$ consisting of a one-body part and a two-body interaction
\begin{eqnarray}
H=\sum_{\alpha\alpha'}\langle \alpha|t|\alpha'\rangle a^\dag_\alpha a_{\alpha'}
+\frac{1}{2}\sum_{\alpha\beta\alpha'\beta'}\langle\alpha\beta|v|\alpha'\beta'\rangle
a^\dag_{\alpha}a^\dag_\beta a_{\beta'}a_{\alpha'},
\nonumber \\
\label{totalH}
\end{eqnarray}
where $a^\dag_\alpha$ and $a_\alpha$ are the creation and annihilation operators of a particle at
a time-independent single-particle state $\alpha$.
\subsection{Time-dependent density-matrix theory and ground state}
The formulation of TDDM has been presented in Refs. \cite{WC,GT,hubbard}. To be self-contained, we briefly explain it below.
TDDM gives the coupled equations of motion for the one-body density matrix (the occupation matrix) $n_{\alpha\alpha'}$
and the correlated part of the two-body density matrix $C_{\alpha\beta\alpha'\beta'}$.
These matrices are defined as
\begin{eqnarray}
n_{\alpha\alpha'}(t)&=&\langle\Phi(t)|a^\dag_{\alpha'} a_\alpha|\Phi(t)\rangle,
\\
C_{\alpha\beta\alpha'\beta'}(t)&=&\rho_{\alpha\beta\alpha'\beta'}(t)
-(n_{\alpha\alpha'}(t)n_{\beta\beta'}(t)-n_{\alpha\beta'}(t)n_{\beta\alpha'}(t)) ,
 \label{rho2}
\end{eqnarray}
where $|\Phi(t)\rangle$ is the time-dependent total wavefunction
$|\Phi(t)\rangle=\exp[-iHt] |\Phi(t=0)\rangle$ and $\rho_{\alpha\beta\alpha'\beta'}$ is the two-body density matrix (
$\rho_{\alpha\beta\alpha'\beta'}(t)=\langle\Phi(t)|a^\dag_{\alpha'}a^\dag_{\beta'}a_{\beta}a_{\alpha}|\Phi(t)\rangle$).
We use units such that $\hbar=1$.
The equations of motion for $n_{\alpha\alpha'}$ and $C_{\alpha\beta\alpha'\beta'}$ are derived from
\begin{eqnarray}
i \dot{n}_{\alpha\alpha'}&=&\langle\Phi(t)|[a^\dag_{\alpha'}a_{\alpha},H]|\Phi(t)\rangle
\label{n0}
\\
i\dot{\rho}_{\alpha\beta\alpha'\beta'}&=&\langle\Phi(t)|[a^\dag_{\alpha'}a^\dag_{\beta'}
 a_{\beta}a_{\alpha},H]|\Phi(t)\rangle,
\label{C0} 
\end{eqnarray}
and are written as
\begin{eqnarray}
i \dot{n}_{\alpha\alpha'}&=&
\sum_{\lambda}(\epsilon_{\alpha\lambda}{n}_{\lambda\alpha'}-{n}_{\alpha\lambda}\epsilon_{\lambda\alpha'})
\nonumber \\
&+&\sum_{\lambda_1\lambda_2\lambda_3}
[\langle\alpha\lambda_1|v|\lambda_2\lambda_3\rangle C_{\lambda_2\lambda_3\alpha'\lambda_1}
\nonumber \\
&-&C_{\alpha\lambda_1\lambda_2\lambda_3}\langle\lambda_2\lambda_3|v|\alpha'\lambda_1\rangle],
\label{n}
\end{eqnarray}
\begin{eqnarray}
i\dot{C}_{\alpha\beta\alpha'\beta'}&=&
\sum_{\lambda}(\epsilon_{\alpha\lambda}{C}_{\lambda\beta\alpha'\beta'}
+\epsilon_{\beta\lambda}{C}_{\alpha\lambda\alpha'\beta'}
\nonumber \\
&-&\epsilon_{\lambda\alpha'}{C}_{\alpha\beta\lambda\beta'}
-\epsilon_{\lambda\beta'}{C}_{\alpha\beta\alpha'\lambda})
\nonumber \\
&+&B_{\alpha\beta\alpha'\beta'}+P_{\alpha\beta\alpha'\beta'}+H_{\alpha\beta\alpha'\beta'}+T_{\alpha\beta\alpha'\beta'},
\label{N3C2}
\end{eqnarray}
where $\epsilon_{\alpha\alpha'}$ is given by
\begin{eqnarray}
\epsilon_{\alpha\alpha'}=\langle \alpha|t|\alpha'\rangle
+\sum_{\lambda_1\lambda_2}
\langle\alpha\lambda_1|v|\alpha'\lambda_2\rangle_A 
n_{\lambda_2\lambda_1}
\label{hf}.
\end{eqnarray}
Here the subscript $A$ means that the corresponding matrix is antisymmetrized. 
The matrix $B_{\alpha\beta\alpha'\beta'}$ in Eq. (\ref{N3C2}) does not contain $C_{\alpha\beta\alpha'\beta'}$ and describes 2p-2h and 2h-2p
excitations, while $P_{\alpha\beta\alpha'\beta'}$ and $H_{\alpha\beta\alpha'\beta'}$
contain $C_{\alpha\beta\alpha'\beta'}$ and describe
p-p (and h-h) and p-h
correlations to infinite order, respectively \cite{GT}. These matrices are given in Ref. \cite{GT}.
The matrix $T_{\alpha\beta\alpha'\beta'}$ is given by
\begin{eqnarray}
T_{\alpha\beta\alpha'\beta'}&=&\sum_{\lambda_1\lambda_2\lambda_3}
[\langle\alpha\lambda_1|v|\lambda_2\lambda_3\rangle C_{\lambda_2\lambda_3\beta\alpha'\lambda_1\beta'}
\nonumber \\
&+&\langle\lambda_1\beta|v|\lambda_2\lambda_3\rangle C_{\lambda_2\lambda_3\alpha\alpha'\lambda_1\beta'}
\nonumber \\
&-&\langle\lambda_1\lambda_2|v|\alpha'\lambda_3\rangle C_{\alpha\lambda_3\beta\lambda_1\lambda_2\beta'}
\nonumber \\
&-&\langle\lambda_1\lambda_2|v|\lambda_3\beta'\rangle C_{\alpha\lambda_3\beta\lambda_1\lambda_2\alpha'}],
\label{T-term}
\end{eqnarray}
where $C_{\alpha\beta\gamma\alpha'\beta'\gamma'}$ is the correlated part of a three-body density-matrix, which is
given by
\begin{eqnarray}
C_{\alpha\beta\gamma\alpha'\beta'\gamma'}=\langle\Phi(t)|a^\dag_{\alpha'}a^\dag_{\beta'}a^\dag_{\gamma'}a_{\gamma}a_{\beta}a_{\alpha}|\Phi(t)\rangle
-{\cal A}(n_{\alpha\alpha'}\rho_{\beta\gamma\beta'\gamma'}).
\end{eqnarray}
Here, ${\cal A}$ is an operator which properly antisymmetrizes $n_{\alpha\alpha'}\rho_{\beta\gamma\beta'\gamma'}$ under the exchange of the single-particle indices
such as $\alpha\leftrightarrow\beta$ and $\alpha'\leftrightarrow\beta'$.
The three-body correlation matrix is neglected in the original version of TDDM \cite{WC,GT}. 
Instead of neglecting $C_{\alpha\beta\gamma\alpha'\beta'\gamma'}$ 
we use the approximation
\begin{eqnarray}
C_{\rm p_1p_2h_1p_3p_4h_2}&\approx&\frac{1}{\cal N}\sum_{\rm h}C_{\rm hh_1p_3p_4}C_{\rm p_1p_2h_2h},
\label{purt1}\\
C_{\rm p_1h_1h_2p_2h_3h_4}&\approx&\frac{1}{\cal N}\sum_{\rm p}C_{\rm h_1h_2p_2p}C_{\rm p_1ph_3h_4},
\label{purt2}
\end{eqnarray}
where ${\rm p}$ and ${\rm h}$ refer to particle and hole states, respectively,
and ${\cal N}$ is given by
\begin{eqnarray}
{\cal N}=1+\frac{1}{4}\sum_{\rm pp'hh'}C_{\rm pp'hh'}C_{\rm hh'pp'}.
\label{norm}
\end{eqnarray}
The above expression can be derived by perturbative consideration \cite{hubbard} using the coupled-cluster
doubles wavefunction.
Equations (\ref{n}) and (\ref{N3C2}) satisfy the conservation laws of the total energy and the
total number of particles \cite{WC,GT}.
The ground state in TDDM is given as a stationary solution of the time-dependent equations 
(Eqs. (\ref{n}) and (\ref{N3C2})). In this work
we use the following adiabatic method to obtain a nearly stationary
solution \cite{toh94,adiabatic1}: Starting from the Hartree-Fock (HF) configuration,
we solve Eqs. (\ref{n}) and (\ref{N3C2}) gradually increasing the strength of the residual interaction such as  
$v({\bm r}-{\bf r'})\times t/T$. This method is motivated by the Gell-Mann-Low theorem \cite{gell}
and has often been used to obtain approximate ground states \cite{lacroix,pfitz}. To suppress oscillating components which come from the mixing
of excited states, we must take large $T$.

\subsection{Extended RPA}

The ERPA equation can be derived by either taking the small amplitude limit of the TDDM equations or using the equation of motion approach
\cite{rowe1}. It is written in matrix form
for the one-body and two-body  amplitudes $x^\mu_{\alpha\alpha'}$ and $X^\mu_{\alpha\beta\alpha'\beta'}$ \cite{Takahara,ts2008}
\begin{eqnarray}
\left(
\begin{array}{cc}
A&B\\
C&D
\end{array}
\right)\left(
\begin{array}{c}
{x}^\mu\\
{X}^\mu
\end{array}
\right)
=\omega_\mu
\left(
\begin{array}{cc}
S_{1}&T_{1}\\
T_{2}&S_{2}
\end{array}
\right)
\left(
\begin{array}{c}
{x}^\mu\\
{X}^\mu
\end{array}
\right),
\label{ERPA1}
\end{eqnarray}
where $\omega_\mu$ is the excitation energy of an excited state $\mu$.
The matrices $A$, $B$, $C$ and $D$ are given by
\begin{eqnarray}
A(\alpha\alpha':\lambda\lambda')&=&\langle\Phi_0|[[a^\dag_{\alpha'}a_\alpha,H],a^\dag_\lambda a_{\lambda'}]|\Phi_0\rangle,
\\
B(\alpha\alpha':\lambda_1\lambda_2\lambda_1'\lambda_2')&=&C^\dag=\langle\Phi_0|[[a^\dag_{\alpha'}a_\alpha,H],a^\dag_{\lambda_1}a^\dag_{\lambda_2} a_{\lambda_2'}a_{\lambda_1'}]|\Phi_0\rangle,
\nonumber \\
D(\alpha\beta\alpha'\beta':\lambda_1\lambda_2\lambda_1'\lambda_2')&=&
\langle\Phi_0|[[a^\dag_{\alpha'}a^\dag_{\beta'}a_\beta a_\alpha,H],a^\dag_{\lambda_1}a^\dag_{\lambda_2} a_{\lambda_2'}a_{\lambda_1'}]|\Phi_0\rangle,
\label{H-matrix}
\end{eqnarray}
where $|\Phi_0\rangle$ is the ground state in ERPA but approximated by that in TDDM. The norm matrices $S_1$, $T_1$, $T_2$ and $S_2$ are given by
\begin{eqnarray}
S_1(\alpha\alpha':\lambda\lambda')&=&\langle\Phi_0|[a^\dag_{\alpha'}a_\alpha,a^\dag_\lambda a_{\lambda'}]|\Phi_0\rangle,
\\
T_1(\alpha\alpha':\lambda_1\lambda_2\lambda_1'\lambda_2')&=&T_2^\dag=\langle\Phi_0|[a^\dag_{\alpha'}a_\alpha,a^\dag_{\lambda_1}a^\dag_{\lambda_2} a_{\lambda_2'}a_{\lambda_1'}]|\Phi_0\rangle,
\nonumber \\
S_2(\alpha\beta\alpha'\beta':\lambda_1\lambda_2\lambda_1'\lambda_2')&=&
\langle\Phi_0|[a^\dag_{\alpha'}a^\dag_{\beta'}a_\beta a_\alpha,a^\dag_{\lambda_1}a^\dag_{\lambda_2} a_{\lambda_2'}a_{\lambda_1'}]|\Phi_0\rangle.
\label{N-matrix}
\end{eqnarray}
The effects of ground-state correlations are included in the above matrices through the occupation matrix and the 
two-body and three-body correlation matrices. 
Although 
the relations $A=A^\dag$ and $C^\dag=B$ hold, $D=D^\dag$ is not satisfied 
because a condition from the Jacobi identity\cite{ts2008}, which is given by
\begin{eqnarray}
\langle\Phi_0|[[a^\dag_{\alpha'} a^\dag_{\beta'}a_\beta a_\alpha,H], a^\dag_{\lambda_1} a^\dag_{\lambda_2} a_{\lambda_2'} a_{\lambda_1'}]|\Phi_0\rangle
&-&
\langle\Phi_0|[[a^\dag_{\lambda_1} a^\dag_{\lambda_2} a_{\lambda_2'} a_{\lambda_1'},H], a^\dag_{\alpha'} a^\dag_{\beta'}a_\beta a_\alpha]|\Phi_0\rangle
\nonumber \\
&=&
-\langle\Phi_0|[H,[a^\dag_{\alpha'} a^\dag_{\beta'}a_\beta a_\alpha,a^\dag_{\lambda_1} a^\dag_{\lambda_2} a_{\lambda_2'} a_{\lambda_1'}]]|\Phi_0\rangle,
\nonumber \\
&=&0
\label{jacobi}
\end{eqnarray}
is not fulfilled:
Since $[a^\dag_{\alpha'} a^\dag_{\beta'}a_\beta a_\alpha,a^\dag_{\lambda_1} a^\dag_{\lambda_2} a_{\lambda_2'} a_{\lambda_1'}]$ in 
the above equation includes three-body
operators, the stationary condition for the three-body density matrix
$\langle\Phi_0|[H,a^\dag_\alpha a^\dag_\beta a^\dag_\gamma a_{\gamma'} a_{\beta'} a_{\alpha'}]|\Phi_0\rangle=0$ is required but it is 
not satisfied by the approximate three-body correlation matrix given by Eqs. (\ref{purt1}) and (\ref{purt2}).
Thus the Hamiltonian matrix of Eq. (\ref{ERPA1}) is not Hermitian.
As a consequence some eigenstates of Eq. (\ref{ERPA1}) can have complex eigenvalues and negative transition strength. 
It does not cause any serious problems in the applications shown below as long as the interaction is not so strong
and the approximation for the three-body correlation matrix is meaningful.

In the following we explain how small-norm components become non-negligible.
The matrix $S_1$ is given by 
\begin{eqnarray}
S_{1}(\alpha\alpha':\lambda\lambda')=(n_{\alpha'\alpha'}
-n_{\alpha\alpha})\delta_{\alpha\lambda}\delta_{\alpha'\lambda'}.
\label{snorm}
\end{eqnarray}
In HF where $n_{\alpha\alpha}=0$ or 1 the diagonal elements of $S_1$ are 1 for the p-h component and $-1$ for the h-p component. When the single-particle states are fractionally occupied, the p-p and h-h elements of $S_1$
become non-vanishing though they are small. As a consequence the contribution of the p-p and h-h amplitudes to Eq. (\ref{ERPA1}) become non-negligible. 
The matrix $S_2$ includes $n_{\alpha\alpha'}$, $C_{\alpha\beta\alpha'\beta'}$ and $C_{\alpha\beta\gamma\alpha'\beta'\gamma'}$.
When $C_{\alpha\beta\alpha'\beta'}$ and $C_{\alpha\beta\gamma\alpha'\beta'\gamma'}$ are neglected for simplicity, 
the diagonal element of $S_2$ is given by \cite{ts2008}
\begin{eqnarray}
S_2(\alpha\beta\alpha'\beta':\alpha\beta\alpha'\beta')&=&(1-n_\alpha)(1-n_\beta)n_{\alpha'}n_{\beta'}
\nonumber \\
&-&n_\alpha n_\beta (1-n_{\alpha'})(1-n_{\beta'}),
\label{S2}
\end{eqnarray}
where we assume that $n_{\alpha\alpha'}=\delta_{\alpha\alpha'}n_\alpha$.
In HF $S_2$ is not vanishing only for the 2p-2h and 2h-2p configurations:
$S_2$ is 1 ($-1$) for the 2p-2h (2h-2p) configurations. When the single-particle states are fractionally occupied, other two-body configurations
such as ph-ph, 2p-2p and 2h-2h configurations
have non-vanishing values of $S_2$ though they are small, and can contribute to Eq. (\ref{ERPA1}). 
The matrices $T_1$ and $T_2$, which can couple the p-p and h-h amplitudes to the 2p-2h and 2h-2p amplitudes and the p-h and h-p amplitudes to the ph-ph amplitudes, are given by $C_{\alpha\beta\alpha'\beta'}$. 

To explain the relation of ERPA to rRPA and SCRPA, we explicitly show
the matrix $A$:
\begin{eqnarray}
A(\alpha\alpha':\lambda\lambda')&=&
(\epsilon_\alpha-\epsilon_{\alpha'})
(n_{\alpha'\alpha'}
-n_{\alpha\alpha})\delta_{\alpha\lambda}\delta_{\alpha'\lambda'}
\nonumber \\
&+&(n_{\alpha'\alpha'}-n_{\alpha\alpha})(n_{\lambda'\lambda'}-n_{\lambda\lambda})\langle\alpha\lambda'|v|\alpha'\lambda\rangle_A
\nonumber \\
&-&\delta_{\alpha'\lambda'}\sum_{\gamma\gamma'\gamma''}\langle\alpha\gamma|v|\gamma'\gamma''\rangle
C_{\gamma'\gamma''\lambda\gamma}
\nonumber \\
&-&\delta_{\alpha\lambda}\sum_{\gamma\gamma'\gamma''}\langle\gamma\gamma'|v|\alpha'\gamma''\rangle
C_{\lambda'\gamma''\gamma\gamma'}
\nonumber \\
&+&\sum_{\gamma\gamma'}(\langle\alpha\gamma|v|\lambda\gamma'\rangle_A
C_{\lambda'\gamma'\alpha'\gamma}
+\langle\lambda'\gamma|v|\alpha'\gamma'\rangle_A
C_{\alpha\gamma'\lambda\gamma})
\nonumber \\
&-&\sum_{\gamma\gamma'}(\langle\alpha\lambda'|v|\gamma\gamma'\rangle
C_{\gamma\gamma'\alpha'\lambda}
+\langle\gamma\gamma'|v|\alpha'\lambda\rangle
C_{\alpha\lambda'\gamma\gamma'}),
\label{A-term}
\end{eqnarray}
where $\epsilon_{\alpha\alpha'}$ and $n_{\alpha\alpha'}$ are assumed to be diagonal for simplicity.
The first two terms on the right-hand side of Eq. (\ref{A-term}) are of the same form as those in the RPA equation, 
the next two terms with $C_{\alpha\beta\alpha'\beta'}$ describe the 
self-energies of the $\alpha-\alpha'$ configurations due to ground-state correlations \cite{Janssen}, and
the last four terms with $C_{\alpha\beta\alpha'\beta'}$ are
interpreted as the modification of the interaction \cite{Janssen}.
The one-body sector of Eq. (\ref{ERPA1}) $Ax^\mu=\omega_\mu S_{1}x^\mu$ is formally the same as the equation
in SCRPA \cite{scrpa1,scrpa2,Janssen}.
In rRPA $C_{\alpha\beta\alpha'\beta'}$ is neglected. 
In rRPA $n_{\alpha\alpha}$ is self-consistently calculated from the p-h and h-p amplitudes, and both $n_{\alpha\alpha}$
and $C_{\alpha\beta\alpha'\beta'}$ are self-consistently obtained from the p-h and h-p amplitudes in SCRPA.
We perform calculations using rRPA-like and SCRPA-like formulations where $n_{\alpha\alpha}$ and $C_{\alpha\beta\alpha'\beta'}$ 
are not given by the p-h and h-p amplitudes but replaced by those of the TDDM ground state.

\begin{figure}[h]
\begin{center} 
\includegraphics[height=8cm]{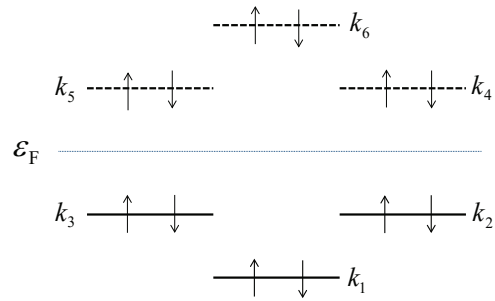}
\end{center}
\caption{Level scheme of the the one-dimensional (1D) Hubbard model in momentum space. Each level can be occupied by spin up ($\uparrow$) and down ($\downarrow$) particles.}
\label{diag}
\end{figure}
\section{Results}
To test our approach for a case where a few h and p states are involved,
we chose the one-dimensional (1D) Hubbard model with periodic boundary conditions
and compare with the results in exact diagonalization approach (EDA).
In momentum space the Hamiltonian is given by \cite{Jemai}
\begin{eqnarray}
H&=&\sum_{{ k},\sigma}\epsilon_{k}a^\dag_{{ k},\sigma}a_{{ k},\sigma}
\nonumber \\
&+&\frac{U}{2N}\sum_{{ k},{ p},{ q},\sigma}
a^\dag_{{ k},\sigma}a_{{ k}+{ q},\sigma}a^\dag_{{ p},-\sigma}a_{{ p}-{ q},-\sigma},
\end{eqnarray}
where 
$U$ is the on-site Coulomb matrix element, $\sigma$ spin projection and 
the single-particle energies are given by $\epsilon_\alpha=-2t\cos k_\alpha$
with the nearest-neighbor hopping potential $t$.
We consider the case of the six sites at half filling.
In the first Brillouin zone $-\pi\le k <\pi$ there are the following wave numbers
\begin{eqnarray}
k_1&=&0,~~~k_2=\frac{\pi}{3},~~~k_3=-\frac{\pi}{3},
\nonumber \\
k_4&=&\frac{2\pi}{3},~~~k_5=-\frac{2\pi}{3}.~~~k_6=-\pi.
\end{eqnarray}
The single-particle energies are $\epsilon_1=-2t$, $\epsilon_2=\epsilon_3=-t$, 
$\epsilon_4=\epsilon_5=t$ and $\epsilon_6=2t$. In HF the lowest 6 states are fully occupied as shown in Fig. \ref{diag}.

\begin{figure}
\begin{center} 
\includegraphics[height=6cm]{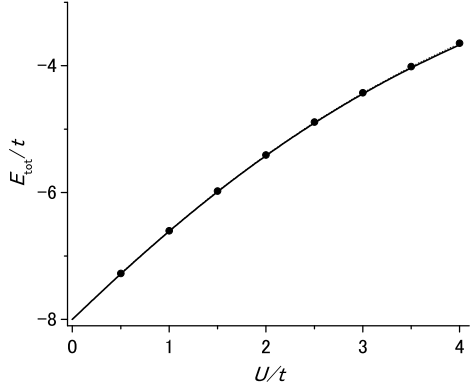}
\end{center}
\caption{Ground-state energy $E_{\rm tot}$ calculated in TDDM (circles)
as a function of $U/t$ for $N=6$ with half-filling. The results in EDA are shown with the solid line.}
\label{etot}
\end{figure}
\begin{figure}
\begin{center} 
\includegraphics[height=6cm]{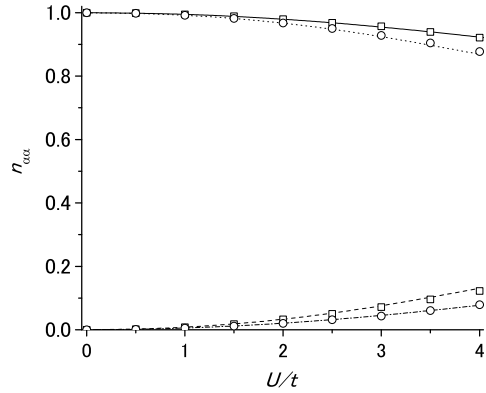}
\end{center}
\caption{Occupation probability of each single-particle state calculated in TDDM (circles and squares) as a function of $U/t$.
The results in EDA are shown with the solid, dotted, dashed and dot-dashed lines.}
\label{hubn}
\end{figure}

\subsection{Ground state}
The ground state energy calculated in TDDM (circles) using the adiabatic method is shown in Fig. \ref{etot} as a function of $U/t$.
The elements of $C_{\alpha\beta\alpha'\beta'}$ which have odd number of p and h states such as $C_{\rm phh'h''}$ and $C_{\rm hpp'p''}$ are neglected 
to reduce the size of the two-body correlation matrix. Their contributions are small.
The results in EDA are given with the solid line.
The TDDM results are in good agreement with the exact values.
The occupation probabilities of the four single-particle states calculated in TDDM (circles and squares)
are compared with the results in EDA (solid,  dotted, dashed and dot-dashed lines) in Fig. \ref{hubn} as functions of $U/t$.
The agreement of the TDDM results with the exact values is also reasonably good. 
 At $U/t=4$ the occupation probabilities deviate more than 10 \%
from the HF values ($n_{\alpha\alpha}=1~{\rm or}~0$).

\begin{figure}[h]
\begin{center} 
\includegraphics[height=6cm]{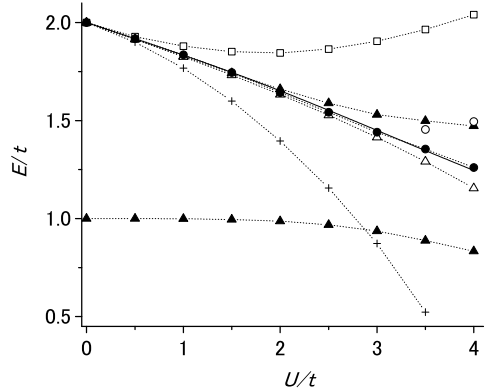}
\end{center}
\caption{Excitation energy of the spin mode with momentum transfer $q=\pi/3$ as a function of $U/t$ calculated in ERPA (filled circles). The results in RPA, rRPA-like and SCRPA-like approaches and SRPA are
shown with the open triangles, filled triangles, squares and crosses, respectively. The results in EDA are depicted with the solid line.
The open circles at $U/t=3.5$ and 4 indicate the ERPA results with the original three-body correlation matrix.}
\label{e2}
\end{figure}

\subsection{Spin mode with momentum transfer $|q|=\pi/3$ \label{second}}
Since
the h-h and p-p transitions such as $k_1\rightarrow k_2$ and  $k_6\rightarrow k_5$ can contribute
when the single-particle states are partially occupied, we first consider the spin mode with the momentum transfer $q=\pi/3$ which can be excited mainly by the one-body operator 
$a^\dag_{k_4\uparrow}a_{k_2\uparrow}-a^\dag_{k_4\downarrow}a_{k_2\downarrow}$.  
The excitation energies calculated in RPA (open triangles), rRPA-like (triangles) and SCRPA-like (squares) formulations
are shown in Fig. \ref{e2} as functions of $U/t$. These approaches do not have the coupling to the two-body amplitudes. The results in EDA are given with the solid line.
Since $U$ is a repulsive interaction,
spin modes where the single-particle transitions between spin-down states destructively interfere with those between spin-up states
become soft with increasing $U$.
The results in RPA are in reasonable agreement with the exact solutions. 
In the rRPA-like calculations the two states appear below $E/t<2$. The lower state at $E/t\approx 1$ mainly consists of the p-p and h-h transitions and the higher state consists of the p-h and h-p
components. The transition strength carried by the lower state increases with increasing $U$. 
Thus in the rRPA-like approach the excited states consisting of the p-p and h-h components appear as if they are physical ones.
This indicates that the inclusion of the ground-state correlation effects only through the fractional occupation of the single-particle states is not appropriate.  
In the SCRPA-like calculations the states mainly consisting of the p-p and h-h components move to the high energy region ($E/t>10$) due to the terms in Eq. (\ref{A-term}) with
$C_{\alpha\beta\alpha'\beta'}$. This is because these terms are divided by the small values $n_{\rm pp}-n_{\rm p'p'}$ or $n_{\rm hh}-n_{\rm h'h'}$ when $Ax^\mu=S_1x^\mu$ is diagonalized.
Thus in the SCRPA-like approach the p-p and h-h components are mixed with the p-h and h-p components but their main components are energetically separated from the low-lying state.
However, the excitation energies calculated in the SCRPA-like approach are significantly larger than the exact values, suggesting the importance of the coupling to the two-body amplitudes.

Now we discuss the coupling to the two-body amplitudes. Since the coupling to the 2p-2h amplitudes is large, the results in SRPA (crosses) deviate strongly from the RPA values with increasing $U$ and SRPA collapses at $U/t=3.9$ for this
spin mode. 
The results in ERPA are shown with the filled circles. 
For this spin mode
the three-body correlation matrix used in Eq.(\ref{ERPA1}) at $U/t=3.5$ and 4 is reduced by 16 \% 
from the value given by Eqs.(\ref{purt1}) and (\ref{purt2}). 
The original three-body correlation matrix gives the results shown with the open circles at $U/t=3.5$ and 4.
The approximation Eqs.(\ref{purt1})
and (\ref{purt2}) seems to overestimate the three-body correlation matrix for large $U$.
For $U/t\le 3$ the excitation energies in ERPA are little affected by such 16 \% reduction of the three-body correlation matrix. 
The coupling to the two-body amplitudes plays a role in bringing down the results of the SCRPA-like approach.
The 2p-2h and 2h-2p amplitudes also play a dominant role in ERPA as in SRPA:
The 2p-2h and 2h-2p amplitudes alone can bring the spin mode down to the position slightly below the second excited state calculated in the rRPA-like approach.
To clarify the contribution of the p-p and h-h amplitudes, we calculate the transition strength in ERPA at $U/t=4$ in two cases using an operator
$\hat{Q}_q=\sum_k \sigma a^\dag_{k+q,\sigma}a_{k,\sigma}$: In one case ($S_{\rm full}$) 
all components of $x^\mu_{\alpha\alpha}$
are included and in the other case ($S_{\rm ph}$) only the p-h and h-p components of $x^\mu_{\alpha\alpha}$ are included. We found $S_{\rm ph}/S_{\rm full}=0.54$. 
Thus the small-norm components have significant contribution to the transition strength.

\begin{figure}[h]
\begin{center} 
\includegraphics[height=6cm]{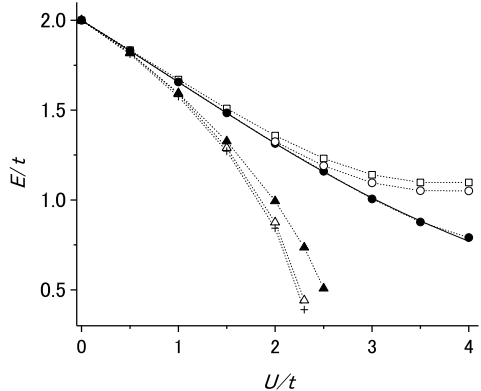}
\end{center}
\caption{Excitation energy of the spin mode with $|q|=\pi$ as a function of $U/t$ calculated in ERPA(filled circles). The results in RPA, rRPA-like and SCRPA-like approaches are
shown with the open triangles, filled triangles and squares, respectively. The crosses depict the results in SRPA. 
The ERPA results calculated using only the 2p-2h and 2h-2p amplitudes are shown with the open circles. The results in EDA are depicted with the solid line.}
\label{e1}
\end{figure}

\subsection{Spin mode with $|q|=\pi$ \label{first}}
Next we consider a spin mode with momentum transfer $|q|=\pi$, which consists of the single-particle transitions from the h states with $k=\pm \pi/3$ to 
the p states with $k=\mp 2\pi/3$. In the Hubbard model the spin mode with $|q|=\pi$ becomes the first excited state for $U>0$. Since only the transitions between 
the single-particle states with $k=\pm 2\pi/3$ and $k=\mp \pi/3$ can make $|q|=\pi$,
there is no contribution to this mode from the p-p and h-h transitions. 
The obtained results are shown in Fig. \ref{e1}.
The spin modes in RPA (open triangles) and SRPA(crosses) collapse beyond $U/t=2.3$. Since the number of the 2p-2h configurations with $|q|=\pi$ is small, the coupling to
the 2h-2h states is small for this mode.
In the rRPA-like approach (filled triangles) which includes the effects of fractional occupation of the single-particles
the point where the spin mode collapses goes slightly up to $U/t=2.6$. This means that fractional occupation of the single-particle states is not sufficient to
suppress excess p-h correlations. In the SCRPA-like approach which includes $n_{\alpha\alpha}$ and $C_{\alpha\beta\alpha'\beta'}$ the spin mode is stable but the excitation energy deviates from the 
exact values with increasing $U$.
In ERPA (filled circles) the coupling to the two-body amplitudes plays a role shifting down the result of the SCRPA-like approach but the 2p-2h and 2h-2p amplitudes 
are not important in contrast to
the case of the spin mode with $|q|=\pi/3$: The ERPA results calculated only with the 2p-2h and 2h-2p amplitudes are shown in Fig. \ref{e1} with the open circles.
Thus the two-body amplitudes which have small eigenvalues of $S_2$ play a role in this spin mode.
\begin{figure}[h]
\begin{center} 
\includegraphics[height=6cm]{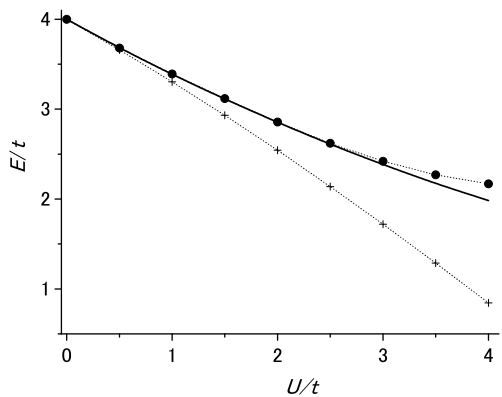}
\end{center}
\caption{Excitation energy of the two-phonon state with $q=0$ calculated in ERPA (circles) as a function of $U/t$. 
The crosses depict the results in SRPA. The results in EDA are
given with the solid line.}
\label{2p}
\end{figure}

\subsection{Two-phonon states with $q=0$}
Finally we consider a state with $q=0$ excited by two-body operators such as 
$a^\dag_{k_4\uparrow}a^\dag_{k_5\downarrow}a_{k_3\downarrow}a_{k_2\uparrow}$.
Since the two-body operator is written as $[a^\dag_{k_4\uparrow}a_{k_2\uparrow}]\times[a^\dag_{k_5\downarrow}a_{k_3\downarrow}]$
or $-[a^\dag_{k_4\uparrow}a_{k_3\downarrow}]\times[a^\dag_{k_5\downarrow}a_{k_2\uparrow}]$,
the state with $q=0$ corresponds to the two-phonon states of the spin modes with $|q|=\pi/3$ and $\pi$ discussed above. 
In HF the matrix elements of $S_2$ for the 2p-2p, 2h-2h and ph-ph amplitudes vanish (see Eq. (\ref{S2})) and SRPA does not have such amplitudes.
In ERPA these amplitudes contribute to Eq. (\ref{ERPA1}). 
Since the two-phonon state with $q=0$ has the same quantum number as the ground state, we neglect
$X^\mu_{\alpha\beta\alpha'\beta'}$ with odd number of p and h states,
to be consistent with the treatment of the ground state. 
The excitation energy of the two-phonon state in ERPA (circles) is shown in Fig. \ref{2p} as a function of $U/t$. The results in EDA are given with the solid line.
The results in ERPA slightly overestimate the excitation energy for large $U$ but are in good agreement with the exact values.
In SRPA the excited energy goes down with increasing $U$ and deviates from the exact values. The difference between the results in ERPA and SRPA demonstrates
the importance of the effects of ground-state correlations. 

\begin{figure}
\begin{center} 
\includegraphics[height=6cm]{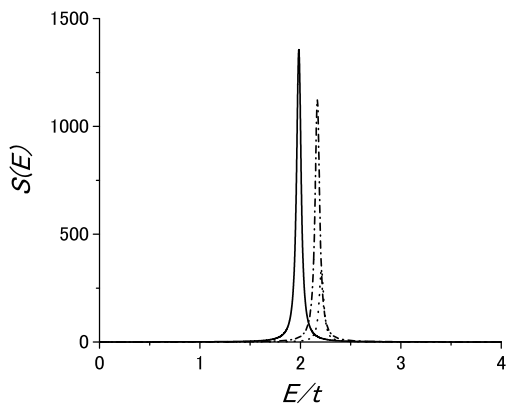}
\end{center}
\caption{Strength distribution calculated in ERPA (dot-dashed line) at $U/t=4$ for the two-phonon state. 
The dotted line depict the ERPA result where only the 2p-2h and 2h-2p components of the two-body amplitude are included as the two-body amplitude. The result in EDA is 
given with the solid line. The distributions are smoothed with width $\Gamma/t=1/20$.}
\label{2pse}
\end{figure}

The strength function calculated in ERPA (solid line) at $U/t=4$ for the operator $\hat{Q}_q^2$ is shown in Fig. \ref{2pse}, where $\hat{Q}_q$ is the excitation operator for the spin mode
with $|q|=\pi$. The dotted line depicts the ERPA result where only the 2p-2h and 2h-2p components are included as the two-body amplitudes. The result in EDA is 
given with the dot-dashed line. The distributions are smoothed with artificial width $\Gamma/t=1/20$. The full ERPA result has the strength comparable to the exact solution, while the ERPA result
calculated with only the 2p-2h and 2h-2p amplitudes has quite small strength. The coupling of the 2p-2h amplitudes to the 2h-2p amplitudes is essential to increase the collectivity of two-phonon
states \cite{toh01}. The ph-ph amplitude plays an important role in such a coupling because the number of the ph-ph configurations is much larger than that of the 2p-2p and 2h-2h configurations.

\section{Summary}
The role of the small-norm amplitudes such as  
the particle-particle and hole-hole components of the one-body amplitudes and the two-body amplitudes other than the two particle - two holes amplitudes were investigated for
the one-dimensional Hubbard model 
using an extended RPA (ERPA) derived from the time-dependent density-matrix theory. 
It was found that these amplitudes cannot be neglected in strongly interacting regions $(U/t>1)$ where the effects of ground-state correlations are significant.
In realistic applications of ERPA, however, we are usually forced to truncate the two-body amplitudes 
because the number of their elements increases rapidly with increasing number of single-particle states.

\appendix{}

\end{document}